\documentstyle[aps,prc,epsfig,preprint]{revtex}
\begin{document}
\draft
\title{Neutral-current neutrino reactions in the supernova environment}

\author{J. M. Sampaio$^1$, K. Langanke$^1$, G. Mart\'{\i}nez-Pinedo$^2$ and D. J. Dean$^3$} 
\address{$^1$Institut for Fysik og Astronomi, {\AA}rhus Universitet, DK-8000 {\AA}rhus C, Denmark\\
$^2$ Departement f\"ur Physik und Astronomie, Universit\"at Basel, Basel, Switzerland\\
$^3$ Physics Division, Oak Ridge National Laboratory, Oak Ridge, TN 37831 USA
}
\date{\today}
\maketitle

\begin{abstract}
We study the neutral-current neutrino scattering for four nuclei in the iron 
region. We evaluate the cross sections for the relevant temperatures during 
the supernova core collapse and derive Gamow-Teller distributions from 
large-scale shell-model calculations. We show that the thermal population 
of the excited states significantly enhances 
the cross sections at low neutrino energies. 
Calculations of the outgoing neutrino spectra indicate 
the prospect of neutrino 
upscattering at finite temperatures. Both 
results are particularly notable in
even-even nuclei.  
\end{abstract}


Neutrino scattering plays a fundamental role in the evolution of the core 
collapse of a massive star towards a supernova explosion. Due to the high 
densities in the star's core, both the transport of electromagnetic radiation 
and electronic heat conduction are very slow compared to the short time scale 
of the collapse  \cite{bethe}. Thus, until the core reaches densities
of $\rho\approx 4\times 10^{11}$ gcm$^{-3}$,  almost all of the 
energy of the collapse is transported by the neutrinos that move out of the 
star essentially without resistance. At higher densities 
the neutrinos can become 
trapped inside the core  due to elastic scattering on nucleons and nuclei. 
Subsequently 
neutrinos lose their energy (downscatter) by inelastic neutrino-electron 
scattering. This process rapidly leads to an equilibrium 
between neutrinos and matter 
\cite{bethe}. The thermalization gradually builds a Fermi distribution of 
neutrinos and helps to maintain a larger lepton fraction inside the core.
 
One might naively expect that this 
straightforward mechanism helps to attain a larger
homologous core that leads to a stronger shock wave 
and less overlying iron for the shock wave to photodisintegrate.
On the other hand,
low-energy neutrinos have a longer mean free path and can therefore diffuse 
more easily out of the core. The net effect of the neutrino downscattering 
depends strongly on the initial conditions of the star and relies on detailed 
neutrino transport calculations that include all relevant 
neutrino reactions. 
Haxton pointed out that neutral-current neutrino scattering on nuclei 
involving the 
excitation of the giant resonances can 
lead to significant neutrino cross 
sections and should also be added in 
supernova simulations \cite{haxton}.
This suggestion was incorporated into a collapse simulation performed by
Bruenn and Haxton \cite{Bruenn91}. They included inelastic
neutrino-nucleus scattering, representing the effects of heavy nuclei
solely by $^{56}$Fe. The relevant cross sections were calculated on the
basis of a strongly truncated shell model for the allowed
transitions and the Goldhaber-Teller model for forbidden transitions. These
calculations demonstrated
that  in later stages of the collapse  inelastic neutrino-nucleus scattering
can compete with  inelastic scattering off electrons.

As mentioned above, neutrino interaction with matter is particularly
important for low neutrino energies $E_\nu$. In this energy range, 
the neutrino-nucleus
rates are found to be smaller than the $\nu+e^-$ rates by $\sim 1-2$
orders of magnitude for $E_\nu \le 5$ MeV and densities in the range
$10^{10}-10^{12}$ g/cm$^3$; however, we note that the inelastic
neutrino-nucleus cross section, obtained for the $^{56}$Fe ground state and
used in \cite{Bruenn91}, underestimates the effects of inelastic
neutrino-nucleus scattering in a supernova environment for two reasons.
First, Gamow-Teller (GT) transitions, which strongly dominate the
inelastic cross sections at low neutrino energies, have to overcome the
gap between the $J=0^+$ ground state (for even-even nuclei like $^{56}$Fe)
and the first excited $J=1^+$ state, which resides at around 3 MeV in
$^{56}$Fe (a typical excitation energy for this state in 
even-even nuclei in the iron group). This introduces a
threshold for inelastic neutrino scattering on the ground state of
even-even nuclei and strongly suppresses the cross sections for
low-energy neutrinos. This strong threshold effect is absent for odd-$A$
and odd-odd nuclei, where the ground state has $J > 0$ and is usually
connected to low-lying states by sizable GT transitions.
Second, inelastic neutrino-nucleus scattering occurs at finite
temperature ($T \gtrsim 0.8$ MeV). Hence excited states in the 
nucleus can be thermally
populated. This effectively removes the threshold effect in even-even
nuclei since the increasing level density with
rising temperature
allows for many
GT transitions. Interestingly, this includes transitions to 
nuclear states at lower excitation energies. 
Such transitions correspond to upscattering of
neutrinos, as their energy in the final state is larger than in the
initial state.

The importance of finite-temperature effects in inelastic
neutrino-nucleus scattering was noted by Fuller and Meyer
\cite{Fuller91} who studied these effects for a few even-even
nuclei (including $^{56}$Fe) on the basis of the independent particle
model; however, this model is not well suited for a quantitative description
at collapse temperatures $(\sim 1$ MeV) as it sets the threshold energy 
for GT transitions effectively by the spin-orbit splitting ($\sim 7$ MeV
for $f_{7/2} \rightarrow f_{5/2}$ in $^{56}$Fe).

The description of inelastic neutrino-nucleus scattering 
at finite temperatures
and low neutrino energies requires a model which reproduces both the
spectroscopy and the GT strength distribution of the nucleus
sufficiently well. For Fe-group nuclei it was recently demonstrated
that modern shell model diagonalization calculations using an
appropriate residual interaction are capable of this task
\cite{Caurier99}. Based on the shell model results of
\cite{Caurier99,Langanke00} we will, in the following, calculate 
inelastic neutrino
cross sections at finite temperatures for the even-even nucleus
$^{56}$Fe, the odd-$A$ nuclei $^{59}$Fe and $^{59}$Co and the odd-odd
nucleus $^{56}$Co. In previous work we have studied finite-temperature
effects on charged-current neutrino reactions, again based on shell
model GT distributions 
\cite{jkg}.

An explicit calculation of the cross section at finite temperature $(T
\gtrsim 1)$ MeV includes too many states to derive the GT strength
distribution for each individual state and is hence unfeasible.
We therefore use the following strategy.
We split the cross section into parts
describing neutrino down-scattering ($E_\nu' \ge E_\nu)$ and
upscattering ($E_\nu' \le E_\nu$), where $E_\nu', E_\nu$ are the
neutrino energies in the final and initial states, respectively. For the
downscattering part we apply the Brink hypothesis \cite{auf} which
states that for a given excited level the GT distribution built on this
state, $S_i (E)$, is the same as for the ground state ($S_0 (E)$), but
shifted by the excitation energy $E_i$, i.e., $S_i(E) = S_0 (E-E_i)$.
The Brink hypothesis was proven valid for the GT resonant states; however,
it can fail for specific low-lying transitions. We consider
important low-lying transitions with large phase space in our
second cross section term. Here we include excited states that are
connected by GT transitions to the ground state or lowest excited states
of the nucleus. These contributions are determined by `inversion' of the
shell-model GT distributions of the low-lying states. Then, assuming that
the inelastic neutrino-nucleus scattering cross section at low energies
($E_\nu \lesssim 15$ MeV) is mainly given by GT transitions, 
our neutral-current neutrino cross section reads:
\begin{equation}
\sigma_{\nu}(E_\nu)=\frac{G_F^2\cos^2\theta_C}{\pi}\left[\sum_f 
E_{\nu^\prime,0f}^2B_{0f}({\rm GT}_0)+\sum_{if} 
E_{\nu^\prime,if}^2B_{if}^{\rm back}({\rm GT}_0)\frac{G_i}{G}\right]\label{eq1}
\end{equation}
where $G_F$ is the Fermi constant, $\theta_C$ the Cabibbo angle, and 
$E_{\nu^\prime,if}$ is the energy of the scattered neutrino, $E_{\nu^\prime,if}
=E_\nu+(E_i-E_f)=E_\nu-q_{0,if}$, and $E_i,E_f$ are the initial and
final nuclear energies.
 The first term arises from the Brink's hypothesis. Under this assumption the 
nuclear transitions are independent of the initial state and, consequently, the
 cross section becomes independent of the temperature. 
For high temperatures, many states
 will contribute in a way that variations in the low-lying transitions tend to
 cancel and Brink's hypothesis becomes a valid assumption (examples supporting
 this reasoning can be found in Ref.~\cite{Langanke00}). The second term 
accounts for the backresonances contribution, where the sum runs over both 
initial (i)  and final states (f). 
The former have a thermal weight of 
$G_i=(2J_i+1)\exp{(-E_i/kT)}$, where $J_i$ is the angular momentum, and 
$G=\sum_i G_i$ is the nuclear partition function. 

All the information about the nuclear structure is comprised in the 
$B_{if}(GT_0)$ coefficients, that define the GT strength 
for the Gamow-Teller operator ${\vec \sigma} t_0$ between an initial and
 final state. These were derived from large-scale shell model calculations 
in the complete $pf$-shell and taking a slightly modified 
version of the KB3 
residual interaction, that corrects for the small overbinding at $N=28$ shell 
closure found in the original KB3 force \cite{Caurier99}.    

\begin{figure}
\begin{center}
\epsfig{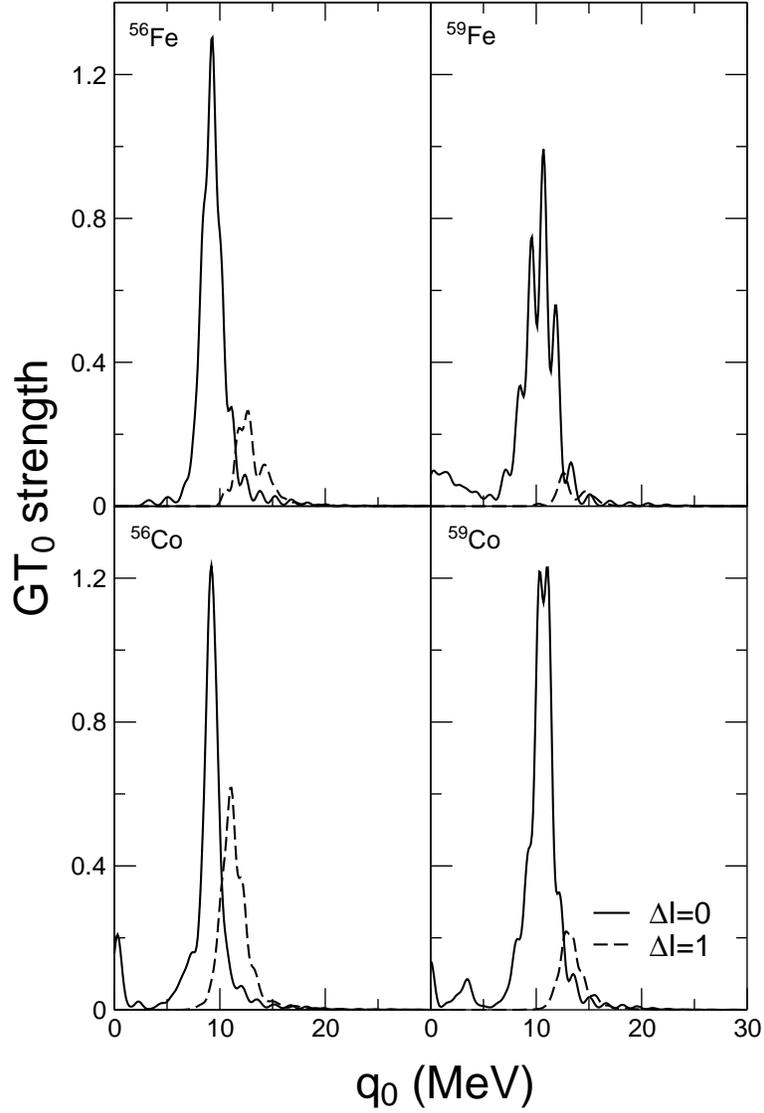}
\end{center}
\caption{Distribution of the GT$_0$ strength built on the 
ground-state for the four nuclei selected for discussion 
throughout this letter. The strength is split into  the two 
isospin components: $\Delta I=0$ (full line) and $\Delta I=1$ (dashed line). 
The energy scale $q_0=E_f-E_i$ refers to the nuclear excitation energy which
is equivalent to the 
neutrino energy transfer.}\label{gt0}
\end{figure}

In Fig. \ref{gt0} we show the GT$_0$ strength built on the ground-state 
of the four 
selected nuclei (with isospin $I$).
One can see that the 
$\Delta I=1$ component has a much lower total strength than the $\Delta I=0$ 
component. This reduction is simply 
due to the geometrical factor that relates the GT$_0$ matrix
element to its reduced matrix element 
and increases 
with larger neutron excess.
The major contribution of the $\Delta I=0$ GT$_0$ strength distribution
is concentrated in a resonant region around $E=10$ MeV for all four
nuclei. (We will refer to this concentration of strength as the GT
resonance.) Hence the position of the GT$_0$ resonance does not depend
on the pairing structure of the nuclear ground state (see also
\cite{Toivanen}). The same behavior was already noticed for the
centroids of the charge-changing GT distributions \cite{Langanke00}.
Remembering that the Brink hypothesis is quite accurate for the GT
resonances one can already conclude here that inelastic neutrino
excitation of the dominant GT$_0$ transitions requires $E_\nu \gtrsim
10$ MeV. Furthermore, a thermal excitation of these states, considered
in the upscattering component, is strongly suppressed at temperatures of
order 1 MeV. Hence, the neutrino cross section for $E_\nu \le 10$ MeV
will be strongly influenced by the rather weak low-energy tail of the
GT$_0$ distribution. In contrast to the resonance peak, 
this tail is quite dependent on the pairing
structure. For odd-$A$ and odd-odd nuclei (with $J \ge 0$) modest GT$_0$
transitions are possible to states at very small excitation energies,
effectively avoiding a threshold for inelastic scattering. This is
clearly different for the even-even nucleus $^{56}$Fe where GT$_0=0$ for
$E \lesssim 3$ MeV. The $\Delta I=1$ component of the GT$_0$
distribution resides at such high excitation energies (centred around 2
MeV higher than the centroids of the $\Delta I=0$ part for the nuclei
studied here), that it cannot be reached by low-energy neutrino
scattering in the upward component and is strongly thermally suppressed
in the downward component of the cross section.

\begin{figure}
\begin{center}
\epsfig{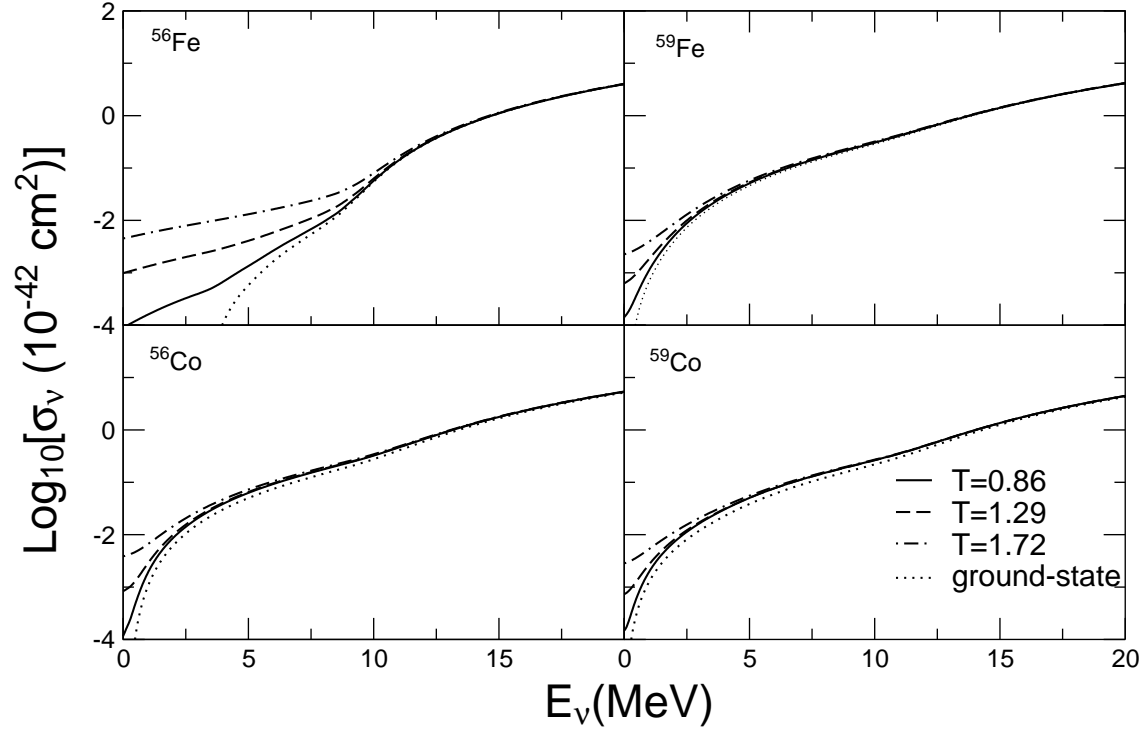}
\end{center}
\caption{Neutrino cross sections from neutrino scattering on nuclei 
at finite temperature. The temperatures are given in MeV. 
The finite temperature results, derived from Eq. \ref{eq1}, 
are compared to the one derived solely 
from the nuclear ground-state.}\label{xsections}
\end{figure}

Fig. \ref{xsections} shows the inelastic neutrino cross sections for the
four nuclei, calculated from
Eq.~\ref{eq1}. We performed the calculations
for the ground state GT$_0$
distribution, which simulates the $T=0$ case, and at 3 different temperatures
$(T=0.86$ MeV, 1.29 MeV, 1.72 MeV), where $T=0.86$ MeV corresponds to
the condition of a presupernova model for a 15$M_\odot$ star ($\rho \sim
10^{10}$ g/cm$^3$ \cite{Heger}). The two other temperatures relate
approximately to neutrino trapping ($T=1.29$ MeV) and thermalization
($T=1.72$ MeV). Our calculations do not include neutrino blocking in the
final state.
For all nuclei we obtain an 
enhancement of the cross sections at finite temperature due to the thermal 
population of the backresonances, i.e., the upwards component. 
Due to the energy gap in the GT$_0$ 
distribution this is particularly pronounced for $^{56}$Fe below 
$E_{\nu}\approx 10$ MeV. For $T=0$ 
the cross section drops 
rapidly to zero  as it approaches the reaction threshold 
($\approx 3$ MeV for $^{56}$Fe). At finite temperature, the gap is then  
filled by thermal population of states with GT$_0$ transitions to the
ground and lowest excited states.
For the odd-A and odd-odd
nuclei there is no  finite threshold at zero temperature and the 
increase of the low-energy cross section with temperature is much 
less dramatic than for
even-even nuclei. Finite temperature effects are unimportant for $E_\nu
\ge 10$ MeV where inelastic  
excitation of the GT$_0$ resonance becomes possible and dominates the
cross sections.

We further note that the cross sections for $^{59}$Fe, $^{59}$Co and
$^{56}$Co are quite similar at finite temperatures indicating the
prospect that inelastic neutrino-nucleus scattering, at least for
odd-$A$ and odd-odd nuclei can be well represented by an `average
nucleus' in collapse simulations. 
We will investigate this point more fully 
in future
research that will also include forbidden transitions. The $^{56}$Fe cross
section is smaller than that of  the other nuclei at all temperatures 
considered here, although this effect is significantly diminished with
increasing temperature.

As Bruenn and Haxton compared the inelastic neutrino-electron and
neutrino-$^{56}$Fe rates, where the latter was evaluated using
shell-model ground-state GT$_0$ distributions 
at low neutrino energies, it is interesting
to compare the increase of the neutrino-nucleus rate quantitatively.
Collapse simulations usually bin the rates in energy intervals of a few
MeV ($\sim 5$ MeV). Thus, we averaged the various inelastic cross
sections between 0 and 5 MeV and observed an increase of the
$^{56}$Fe($\nu,\nu')$ cross section by about a factor 30 as the
temperature is increased 
from $T=0$ to $T=1.29$ MeV. For $T=0$ the averaged cross
section for the other nuclei is nearly 200 times larger than the one for
$^{56}$Fe. At $T=1.29$ MeV the cross section for $^{59}$Fe, $^{59}$Co and
$^{56}$Co are also increased by about $30\%$ due to upscattering
contributions. This implies that the inelastic neutrino-nucleus
scattering rate is larger at low energies than assumed in \cite{Bruenn91}
by more than an order of magnitude. 

During the collapse, inelastic scattering off electrons and nuclei is
most important in the thermalization phase with densities between
$\rho \sim 10^{11}-10^{12}$ g/cm$^3$ and temperatures $T \sim 1-1.75$ MeV.
Under such conditions neutrinos are produced mainly by electron capture
on free protons and hence have energies of order $8-15$ MeV
\cite{Bruenn91}. If we compare the various inelastic neutrino cross
sections, for example at $T=1.29$ MeV, with the $^{56}$Fe $T=0$ cross section,
as used in \cite{Bruenn91}, we find an increase of about a factor 5-6 for
$E_\nu=10$ MeV for the two cobalt isotopes; at $E_\nu=15$ MeV this
increase compared to the $T=0$ $^{56}$Fe cross section is reduced to
about $30\%$. Hence, we conclude that nuclear structure and
temperature effects might still be relevant for 10 MeV neutrinos, but
can be neglected for $E_\nu \gtrsim 15$ MeV.

\begin{figure}
\begin{center}
\epsfig{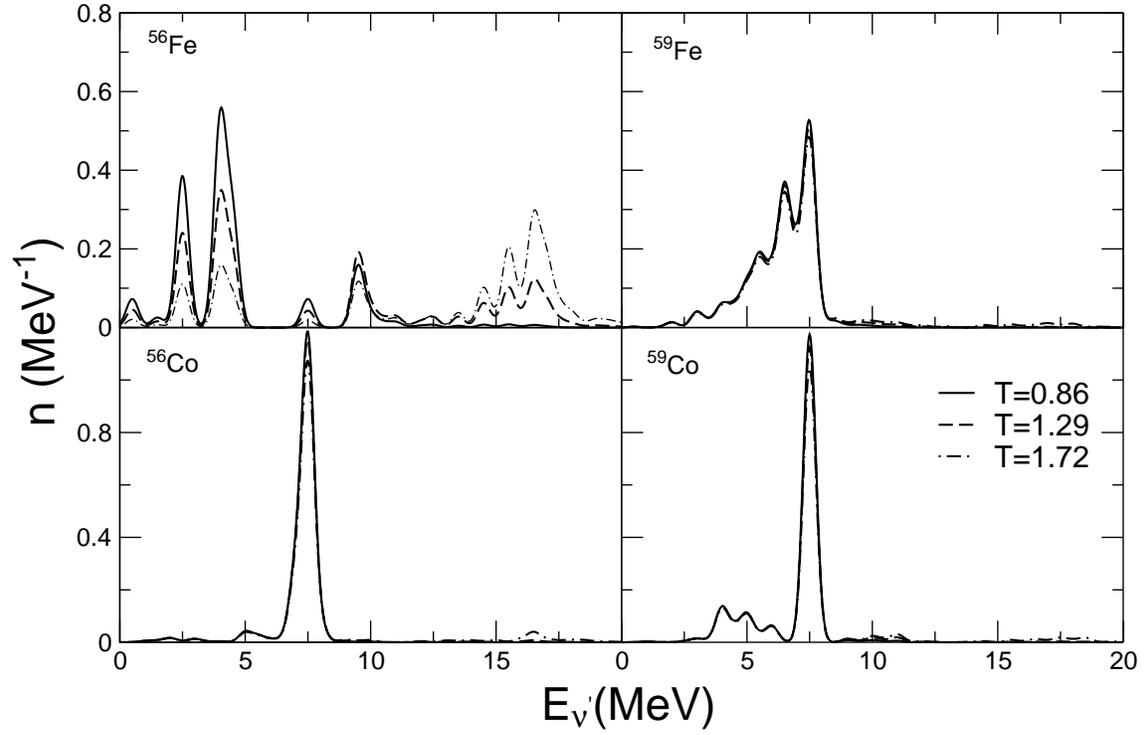}
\end{center}
\caption{Normalized neutrino spectra for inelastic scattering of 
$E_{\nu}=7.5$ MeV neutrinos 
on nuclei at finite temperature. The temperatures are 
given in MeV.}\label{spectra1}
\end{figure}

\begin{figure}
\begin{center}
\epsfig{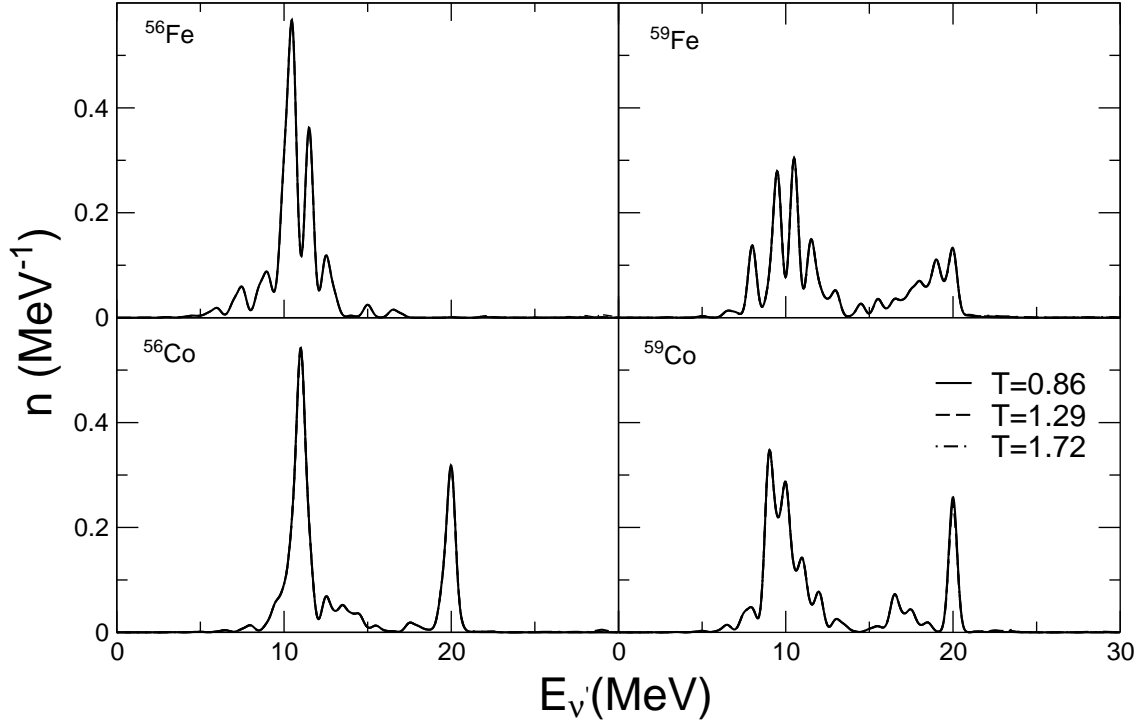}
\end{center}
\caption{Normalized neutrino spectra for inelastic scattering of
$E_{\nu}=20$ MeV neutrinos 
on nuclei at finite temperature. The temperatures are 
given in MeV. Note that the three temperature curves completely 
overlap, as the spectra are dominated by the (temperature-independent)
downscattering components. }\label{spectra2}
\end{figure}

In our model, the temperature-related increase of the cross section is
due to the upscattering contributions when the neutrino gains
energy by nuclear deexcitation. We note that the previously discussed
downscattering contributions in inelastic neutrino scattering increase
the entropy of the collapse environment as they result in nuclear
excitation. In contrast, the upscattering contributions reduce the
entropy. The relative importance of the down- and upscattering
components can also be read directly from the neutrino energy distributions
resulting from the inelastic scattering process. 
Such double-differential distributions
are also of interest for the neutrino transport simulations
\cite{Mezzacappa}. We present two representative neutrino spectra for initial
$E_\nu=7.5$ MeV and 20 MeV neutrinos in Figs. \ref{spectra1} and
\ref{spectra2}, respectively. The spectra are normalized to unity.
Neutrinos with $E_\nu=7.5$ MeV are typical for the early collapse stage,
while neutrino energies of about 20 MeV are encountered during the
thermalization phase at densities around $10^{12}$ g/cm$^3$.
For $E_\nu=7.5$ MeV neutrinos, the upscattering contributions noticeably 
increase the cross sections for $^{56}$Fe. Related to this, we also find a
significant portion of the spectrum at energies $E_\nu' \ge 10$ MeV. With
increasing temperature, the thermal population of the excited states
increases and so do the upscattering contributions to cross sections
and spectra. The sizable bump in the spectrum at around $E_\nu' \sim 17$
MeV corresponds to thermal excitations of the GT$_0$ resonance. 
Except for minor high-energy wings with $E_\nu' \ge 7.5$ MeV, reflecting
upscattering, finite temperature effects are rather unimportant for the
normalized neutrino spectra in inelastic scattering off the other 3 nuclei. 
The spectra for these nuclei, in particular for $^{56,59}$Co, are
dominated by peaks around the incoming neutrino energy ($E_\nu=7.5$ MeV)
implying that excitation and deexcitation of low-lying excited states
dominate the cross sections.

We note that the GT$_0$ resonance
cannot be reached in the downscattering direction for
$E_\nu=7.5$ MeV neutrinos. This is different for $E_\nu=20$ MeV
neutrinos. Consequently the cross section and spectra are entirely
dominated by the downscattering contributions for all four nuclei, as
already reasoned above (Fig. \ref{spectra2}). The noticeable peak in the
spectra around $E_\nu^\prime=10$ MeV correspond to the inelastic excitation
of the GT$_0$ resonance. 	 In a quantitative
calculation of the cross section in this energy regime
forbidden transitions have to be included \cite{Toivanen}; however,
they do not change our conclusion that finite temperature effects
can be neglected.

In summary, we have shown that finite temperature enhances the neutrino 
inelastic scattering cross sections for low-energy neutrinos. The
increase is most significant for even-even nuclei. Our results suggest
that inelastic neutrino-nucleus scattering rates might be comparable to
the inelastic neutrino scattering off electrons at low neutrino
energies. The increase in cross section is related to
deexcitation of thermally populated states and thus reduces the entropy and
scatters the neutrinos up in energy. We also find that thermal effects
will only influence the cross section for neutrino energies which do not
allow an excitation of the GT$_0$ resonance. As these resonances reside
around $E=10$ MeV in $pf$-shell nuclei, thermal effects are restricted
to neutrinos with $E_\nu \le 10$ MeV.

Our calculation also indicates rather mild variations of the inelastic
neutrino cross section for different nuclei at collapse temperatures. 
This implies the
possibility to derive a double-differential cross section for an
`average nucleus' for implementation in collapse simulations, at least
for odd-A and odd-odd nuclei. Such work
is in progress.

\acknowledgements 

Discussions with W.C. Haxton are thankfully appreciated. 
Our work has been supported by the Danish 
Research Council and by the Swiss National Science Foundation. 
J.M.S. acknowledges the financial support of the 
Portuguese Foundation for Science and Technology.  Oak Ridge National
Laboratory is managed by UT-Battelle, LLC.~for the U.S. Department of
Energy under contract No. DE-AC05-00OR22725.

\newpage

\newpage

\newpage

\end{document}